\documentstyle[11pt]{article}
\def\ATMP#1#2#3{{\it Adv. Theor. Math. Phys.} {\bf #1} {(#2)} {#3}}
\def\JHEP#1#2#3{{\it JHEP} {\bf #1} {(#2)} {#3}}
\def\NPB#1#2#3{{\it Nucl. Phys.} {\bf B#1} {(#2)} {#3}}
\def\PLB#1#2#3{{\it Phys. Lett.} {\bf B#1} {(#2)} {#3}}
\def\CQG#1#2#3{{\it Class. Quantum Grav.} {\bf #1} {(#2)} {#3}}
\def\MPLA#1#2#3{{\it Mod. Phys. Lett.} {\bf A#1} {(#2)} {#3}}

\def\PRD#1#2#3{{\it Phys. Rev.} {\bf D#1} {(#2)} {#3}}

\def\AP#1#2#3{{\it Ann. Phys.} {\bf #1} {(#2)} {#3}}

\begin{document}
{}~\hfill hep-th/0002008

\vspace{2cm}

\begin{center}

{\Large\bf
$IIB$ supergravity and various aspects of
light-cone formalism in AdS space-time}
\footnote{
Based on talk given at
International Workshop
"Supersymmetries and Quantum Symmetries",
Dubna, Russia, July 27-31, 1999
}

\vspace{1cm}
R.R. Metsaev

\vspace{1cm}
{\it Department of Theoretical Physics, P.N. Lebedev Physical
Institute, Leninsky prospect 53, 117924, Moscow, Russia}

\vspace{2cm}
{\bf Abstract}
\end{center}

\noindent
{\small
Light-cone gauge manifestly supersymmetric formulation of type
IIB  10-dimensional supergravity in $AdS_5 \times S^5$ background
is discussed. The formulation is given entirely in terms of light-cone
scalar superfield, allowing  us to treat all component fields on
an equal footing. Discrete energy spectrum of field propagating in AdS
space is explained within the framework of light-cone approach.
Light-cone gauge formulation of self-dual fields propagating in AdS space
is developed. An conjectured interrelation of higher spin massless fields
theory in AdS  space-time and superstring theory in Minkowski space is
discussed.  }

\newpage
{\bf
Light-cone gauge formulation of $IIB$ supergravity in
$AdS_5 \times S^5$ background.}
In recent years, due to the conjectured duality
between the string theory and ${\cal N}=4$, $4d$ SYM theory
\cite{mal} (for review see \cite{review})
there has been rapidly growing interest in
strings propagating in AdS space.
Inspired by this conjecture the Green-Schwarz formulation of strings
propagating in $AdS_5\times S^5$ was suggested in \cite{mt1} (for  further
developments see \cite{krr}-\cite{k1}).  Despite considerable efforts
these strings have not yet been quantized\footnote{\scriptsize
Some related interesting discussions are in \cite{Dolan:1999pi}.
Alternative approaches can be found in \cite{alter},\cite{ber}.
Twistor-like formulations are discussed in \cite{twi}.}.
As is well
known, quantization of GS superstrings propagating in flat space is
straightforward only in the light-cone gauge.
Because the string theories are approximated at low energies
by supergravity theories it seems reasonable  that we should first study a
light-cone gauge formulation of supergravity theories in AdS spacetime.
Light-cone description of type $IIB$ supergravity in $AdS_5\times S^5$
background can teach us a lot. For example it can be used to construct
charges of global symmetries so that certain of them have the same
form for both supergravity and superstring.
Keeping in mind extremely important applications to string theory let us
now restrict our discussion to $IIB$ supergravity in $AdS_5\times S^5$.

Our method is conceptually very close to the one
used in \cite{GS5} (see also \cite{GSB}) to find the light-cone form of
$IIB$ supergravity in  flat space and is based essentially on a light-cone
gauge description of field dynamics developed recently in
\cite{metlc}\footnote{\scriptsize
A discussion of $IIB$ supergravity at the level of gauge invariant
equations of motion and actions can be found in \cite{covequmot} and
\cite{covact} respectively.}.
To discuss light-cone formulation
we use the  following  parametrization of $AdS_5\times S^5$ space
$$
ds^2=\frac{1}{z^2}(-dt^2+dx_1^2+dx_2^2+dz^2+ dx_4^2)
+\frac{1}{4}dy_{ij}dy_{ij}^*\,,
$$
introduce
light-cone variables $x^\pm \equiv(x^4  \pm x^0)/\sqrt{2}$,
$x\equiv(x^1+{\rm i}x^2)/\sqrt{2}$, $\bar{x}\equiv x^*$, where $x^0\equiv
t$ and treat $x^{+}$ as evolution parameter.
Here and below we set the radii of  both  $AdS_5$ and $S^5$ equal
to unity\footnote{\scriptsize
The $S^5$ coordinates $y^{ij}$ are subject to the constraints
$y^{ik}y_{kj}=\delta_j^i$,
$y_{ij}=\frac{1}{2}\epsilon_{ijkn}y^{kn}$,
$y_{ij}^*=-y^{ij}$
where indices take the values
$i,j,k,n=1,2,3,4$
The $\epsilon^{ijkn}=\pm 1$ is the Levi-Civita tensor of $su(4)$.}.
The coordinates $y_{ij}$ are related to the  standard $so(6)$
cartesian coordinates $y^M$, $M=1,\ldots,6$, which satisfy the constraint
$y^My^M=1$ through the formula $y_{ij}=\rho_{ij}^M y^M$,
where $\rho_{ij}^M$ are the Clebsh Gordan coefficients of $su(4)$
algebra \cite{GS6}.

Our goal is to find  a realization of $psu(2,2|4)$
superalgebra on the space of IIB supergravity fields propagating in
$AdS_5\times S^5$.  To do that we use light-cone superspace
formalism. First, we introduce light-cone superspace which by definition
is based on position $AdS_5\times S^5$ coordinates $x^\pm$, $x$,
$\bar{x}$, $z$, $y^{ij}$ and Grassmann position coordinates $\theta^i$ and
$\chi^i$ which transform in fundamental ${\bf 4}$ irreps of $su(4)$
algebra. Second,  on this superspace we introduce scalar
superfield $\Phi(x^\pm,x,\bar{x},z,y^{ij},\theta^i,\chi^i)$. In the
remainder of this paper we find it convenient to Fourier transform to
momentum space for all coordinates except for $x^+$, $z$ and $S^5$
coordinates $y^{ij}$.  This implies using $p^+$, $\bar{p}$, $p$,
$\lambda_i$, $\tau_i$ instead of $x^-$, $x$, $\bar{x}$, $\theta^i$,
$\chi^i$ respectively. The $\lambda_i$ and $\tau_i$ transform in
$\bar{{\bf 4}}$ irreps of $su(4)$.
Thus we consider the superfield
$\Phi(x^+,p^+,p,\bar{p},z,y^{ij},\lambda_i,\tau_i)$\footnote{\scriptsize
An expansion of $\Phi$ in powers of Grassmann momenta $\lambda_i$ and
$\tau_i$ can be found in \cite{metiib}.}
which by definition satisfies the following reality constraint
$$
\Phi(x^+,-p, z, y,\lambda,\tau) =p^{+4}\int d^4\lambda^\dagger
d^4\tau^\dagger e^{(\lambda_i\lambda_i^\dagger+\tau_i\tau_i^\dagger)/p^+}
(\Phi(x^+,p,z, y,\lambda,\tau))^\dagger\,.
$$
In  the light-cone formalism the $psu(2,2|4)$ superalgebra has
the generators
\begin{equation}\label{kingen}
P^+,\,P,\,\bar{P},\, J^{+x},\,J^{+\bar x},\, K^+,\, K,\,{\bar K},\,
Q^{+i},\,Q^+_i,\,S^{+i},\,S^+_i,\,
D,\,J^{+-},\,J^{x\bar x}\,,
\end{equation}
which we refer to as kinematical generators and
\begin{equation}\label{dyngen}
P^-,\, J^{-x}\,, J^{-\bar x},\,K^-\,, Q^{-i},\, Q^-_i,\, S^{-i},\,S^-_i
\end{equation}
which we refer to as dynamical generators. The kinematical generators
have positive or zero $J^{+-}$ charges, while dynamical generators have
negative $J^{+-}$ charges\footnote{\scriptsize For $x^+=0$ the kinematical
generators are quadratic in the physical field $\Phi$, while the dynamical
generators receive corrections in interaction theory.  Here
we deal with free fields.}.
At a quadratic level both kinematical and
dynamical generators have the following representation in terms of the
physical light-cone superfield
$$
\hat{G}=\int dp^+d^2p dz dS^5d^4\lambda d^4\tau\, p^+
\Phi(x^+,-p,z,y,-\lambda,-\tau) G\Phi(x^+,p,z,y,\lambda,\tau)\,,
$$
where $G$ are the differential operators acting on $\Phi$.
Thus we should find representation of $psu(2,2|4)$ in terms of
differential operators acting on light-cone superfield $\Phi$.
To simplify expressions let us write down the generators
for $x^+=0$. The kinematical generators are then given by
\begin{equation}\label{kin1}
P=p\,,
\qquad
\bar{P}=\bar{p}\,,
\qquad
P^+=p^+\,,
\qquad
J^{+x}=\partial_pp^+\,,
\qquad
J^{+\bar{x}}=\bar{\partial}_pp^+\,,
\end{equation}
\begin{equation}\label{kin3}
Q^{+i}=p^+\theta^i\,,
\quad
Q_i^+=\lambda_i\,,
\quad
S^+_i=\frac{1}{\sqrt{2}}z\tau_i-\lambda_i\partial_p\,,
\quad
S^{+i}=\frac{1}{\sqrt{2}}zp^+\chi^i+p^+\theta^i\bar{\partial}_p\,,
\end{equation}
where $\partial_p\equiv\partial/\partial \bar{p}$,
$\bar{\partial}_p\equiv\partial/\partial p$.
Dynamical generator $P^-$ is given by
\begin{equation}\label{adspm}
P^-=-\frac{p\bar{p}}{p^+}+\frac{\partial^2_z}{2p^+}
-\frac{1}{2z^2p^+}A\,,
\qquad
A\equiv X-\frac{1}{4}\,,
\end{equation}
where
\begin{equation}\label{ax}
X\equiv l^i{}_j^2+4\tau l \chi +(\chi\tau-2)^2,
\quad
l^i{}_j^2\equiv l^i{}_jl^j{}_i\,,
\quad
\tau l\chi \equiv\tau_i l^i{}_j\chi^i\,,
\qquad
\chi\tau\equiv\chi^i\tau_i\,.
\end{equation}
The $l^i{}_j$ is the orbital part of $su(4)$ angular momentum $J^i{}_j$.
Explicit expressions for the remaining kinematical and dynamical generators
may be found in \cite{metiib}\footnote{\scriptsize For $\lambda_i$,
$\tau_i$ and $\tau^i$, $\chi^i$ we adopt the following anticommutation
rules $\{\theta^i,\lambda_j\}=\delta^i_j$,
$\{\chi^i,\tau_j\}=\delta^i_j$.}.
Making  use of the expression for $P^-$ (\ref{adspm}) we can immediately
write down the light-cone gauge action\footnote{\scriptsize
Since the action is invariant with respect to the  symmetries generated by
$psu(2,2|4)$ superalgebra,  the formalism we discuss is  sometimes referred
to as an off shell light-cone formulation \cite{GSB}.}

$$
S_{l.c.}
=\int\! dx^+dp^+ d^2pdz dS^5 d^4\lambda d^4\tau\,
p^+\Phi(x^+,-p,z,y,-\lambda,-\tau)
({\rm i}\partial^-  + P^-)\Phi(x^+,p,z,y,\lambda,\tau).
$$
Following \cite{metlc} we shall call  the
operator $A$  the $AdS$ mass operator. A few comments are in order.

(i)  The operator $A$ is equal to zero only for massless representations
realized as irreducible representations of conformal algebra
\cite{metsit3},\cite{metlc} which for the case of $AdS_5$
space is the $so(5,2)$ algebra.
Because the operator $X$ (\ref{ax}) has eigenvalues equal to squared
integers in the whole spectrum of compactification of $IIB$ supergravity
on $AdS_5$ (see \cite{metiib}) the operator $A$ is never equal
to zero. This implies that the  scalar fields \cite{krn} as well
as all remaining fields of compactification of $IIB$ supergravity do not
satisfy conformally invariant equations of motion.

(ii) The coordinate $\theta^i$ (or $\lambda_i$) constitutes odd part of
light-cone superspace appropriate to superfield description of light-cone
gauge $N=4$, $4d$ (or $N=1$, $10d$) SYM theory. The translation
generators, Lorentz boosts and Poincar\'e supercharges given in
(\ref{kin1}),(\ref{kin3}) take the same form as in SYM theory. From this
we conclude that one can expect that superstring dynamics in
$AdS_5\times S^5$ could be presented as dynamics of free left and right
movers appropriate to description of open superstrings
supplemented by nonlinear dynamics of remaining degrees of freedom.

(iii) The generator $P^-$ involves terms up to fourth order in
$\chi^i$ and $\tau_i$. The first thing to note is that terms of
fourth order in $\chi^i$ and $\tau_i$ appear trough the number operator
$\chi\tau$. The second point is that terms of fourth order can be excluded
by introducing new superfield\footnote{\scriptsize
We prefer to use superfield $\Phi$ instead of
$\Phi^{new}$ because the $\Phi$ has  conventional canonical dimension.}
$\Phi=z^{\chi\tau-2}\Phi^{new}$.
On the space of $\Phi^{new}$ the hamiltonian $P^-$ takes the form
$$
P^-=-\frac{p\bar{p}}{p^+}+\frac{\partial^2_z}{2p^+}
+\frac{\chi\tau-2}{zp^+}\partial_z
-\frac{1}{2z^2p^+}(
l^i{}_j^2+4\tau l \chi + \chi\tau-\frac{9}{4})
$$
i.e. the terms of fourth order in $\chi^i$ and $\tau_i$ are absent. This
suggests (but does not prove) light-cone gauge string action which does
not involve higher than second order terms in anticommuting
variables\footnote{\scriptsize Such action has been found recently in
Ref.\cite{peslc}.}.

\bigskip
{\bf Discrete energy spectrum of field in AdS space-time.}
As is well known energy spectrum of field propagating in AdS space
takes discrete values (see for review \cite{BRFR}).  This phenomenon is not
immediately visible in light-cone formulation. In this section we would
like to explore how the discrete energy spectrum is obtained within the
framework of the light-cone formulation.

The AdS translation generators $\hat{P}^a$ are expressible
as\footnote{\scriptsize
In this section, in contrast to \cite{metlc}, we
use hermitean $\hat{P}^a$, $P^a$, $K^a$ which are
related with the anithermitean ones of \cite{metlc} as follows
$\hat{P}_{herm}^a=-{\rm i}\hat{P}_{ant herm}^a$,
$P_{herm}^a=-{\rm i}P_{anti herm}^a$,
$K_{herm}^a=+{\rm i}K^a_{anti herm}$.}
$\hat{P}^a=(1/2)P^a+K^a$.  The $\hat{P}^0$ is the energy
operator. In light-cone frame we have\footnote{\scriptsize
We use parametrization of $AdS_d$ space in which
$ds^2=(-dx^{02}+dx_I^2+dx_{d-1}^2)/z^2$.
Light-cone
coordinates in $\pm$ directions are defined as $x^\pm=(x^{d-1}\pm
x^0)/\sqrt{2}$.
From now on we adopt the following  conventions:
$I,J=1,\ldots, d-2$; $i,j,k,l=1,\ldots,d-3$.
$\partial^I=\partial_I\equiv\partial/\partial x^I$,
$\partial^+=\partial_- \equiv \partial/\partial x^-$,
$\partial_{p^+}\equiv\partial/\partial p^+$,
$z\equiv x^{d-2}$.
In momentum representation $\partial^+$ takes the form
$\partial^+={\rm i}p^+$.}
$$
\hat{P}^0=\frac{1}{2\sqrt{2}}(P^+ + 2K^+ - P^- -
2K^-)\,.
$$
Representation of generators $P^a$ and $K^a$ on the space of
physical fields has been found in \cite{metlc}. For the case of
totally symmetric fields this representation takes the
form (for $x^+=0$)
\begin{eqnarray}
&&
P^+ = p^+\,,
\qquad\qquad
K^+ = \frac{1}{2}x_I^2p^+\,, \qquad x_I^2\equiv x^Ix^I
\nonumber\\
&&
P^-=\frac{\partial_I^2}{2p^+}
-\frac{1}{2z^2p^+}(-\frac{1}{2}M_{ij}^2+\frac{(d-4)(d-6)}{4})\,,
\qquad
\partial_I^2\equiv \partial^I\partial^I
\nonumber\\
&&
K^-=\frac{1}{2}x_I^2P^-
-\partial_{p^+}D
+\frac{1}{2p^+}l^{IJ}M^{IJ}
-\frac{x^I}{2zp^+}\{M^{zJ},M^{JI}\}\,,
\nonumber\\
\label{dlc}&&
D=-\partial_{p^+}p^+ + x^I\partial^I+\frac{d-2}{2}\,,
\qquad
M^{IJ} \equiv \alpha^I\bar{\alpha}^J-\alpha^J\bar{\alpha}^I\,.
\end{eqnarray}
where $l^{IJ}\equiv x^I\partial^J-x^J\partial^I$
and we adopt a convention:
$\{a,b\}\equiv ab+ba$.
The above generators act on
physical field whose components are collected in Fock vector
$|\phi\rangle$:
\begin{equation}\label{totsym}
|\phi\rangle=\phi^{I_1\ldots I_s}
\alpha^{I_1}\ldots \alpha^{I_s}|0\rangle\,,
\qquad
\bar{\alpha}^I|0\rangle=0\,,
\qquad
[\bar{\alpha}^I,\alpha^J]=\delta^{IJ}\,,
\end{equation}
where physical traceless tensor field $\phi^{I_1\ldots I_s}$ depends on
$x^+,x^i,z,p^+$.  Using
these expressions we get the following representation for energy operator
(for $x^+=0$)
\begin{equation}\label{hpz}
2\sqrt{2}\hat{P}^0=(1+x_I^2)(p^+ - P^-)
+2\partial_{p^+}D
-\frac{1}{p^+}l^{IJ}M^{IJ}
+\frac{x^I}{zp^+}\{M^{zJ},M^{JI}\}\,.
\end{equation}
It is convenient to make the following transformation of wave function
$$
\phi= U \tilde{\phi}\,,
\qquad
U\equiv (p^+)^{\frac{1}{2}(x^I\partial^I+\frac{d-5}{2})}
$$
and use the formula
\begin{equation}\label{pdutp}
\partial_{p^+}D\phi=U\Bigl(
-p^+\partial_{p^+}^2-\frac{1}{2}\partial_{p^+}
+\frac{1}{4p^+}(x_I^2\partial_J^2-
\frac{1}{2}l_{IJ}^2+\frac{(d-3)(d-5)}{4})\Bigr)\tilde{\phi}.
\end{equation}
Now to simplify our presentation let us restrict ourselves to the case of
four dimensional ($d=4$) AdS space.
For this case because the indices $i,j$ which
label $d-3$ directions take one value $i,j=1$ the spin operator
$M^{ij}$ is equal to zero. Now taking into account that
$M_{IJ}^2=2M_{z1}^2$ and exploiting (\ref{pdutp}) in (\ref{hpz}) we get
the following representation of $\hat{P}^0$ in $\tilde{\phi}$
\begin{equation}\label{p0h1h2}
2\sqrt{2}\hat{P}^0=H_1 + H_2\,,
\end{equation}
where
\begin{eqnarray}
&&
H_1\equiv-\frac{1}{2}\partial_I^2+x_I^2\,,
\\
&&
H_2\equiv-2p^+\partial_{p^+}^2 -\partial_{p^+}
-\frac{1}{2p^+}\Bigl(2(\frac{1}{2}l^{IJ}+M^{IJ})^2+\frac{1}{4}\Bigr)
+p^+\,.
\end{eqnarray}
Because the hamiltonians $H_1$, $H_2$ commute with each other we can
diagonalize them simultaneously. Let us find their eigenvalues.
To this end we introduce complex coordinates $x$, $\bar{x}$ instead of
 $x^1,z$: $x\equiv (x^1+{\rm i}z)/\sqrt{2}$, $\bar{x}\equiv x^*$.
Next we decompose the wave function $|\tilde{\phi}\rangle$ as follows
$$
|\tilde{\phi}\rangle=|\phi_{+s}\rangle+|\phi_{-s}\rangle\,,
$$
where the $|\phi_{\pm s}\rangle$ are eigenvectors of $M^{x\bar{x}}$:
$M^{x\bar{x}}|\phi_{\pm s}\rangle=\pm s|\phi_{\pm s}\rangle$.
The $|\phi_{\pm s}\rangle$ themselves can be decomposed into eigenvectors of
operator $l^{x\bar{x}}$
$$
|\phi_{\pm s}\rangle
=\sum_{m=-\infty}^{\infty}e^{{\rm i}m\varphi}|\phi_{\pm s,m}\rangle\,,
\qquad
l^{x\bar{x}}e^{{\rm i}m\varphi}=me^{{\rm i}m\varphi}\,,
\qquad
\varphi\equiv \hbox{arg}(x^1+{\rm i}z)\,.
$$
On space of $|\phi_{\pm s,m}\rangle$ the hamiltonians $H_1$, $H_2$ take the
form
\begin{eqnarray}
&&
H_1=-\frac{1}{2}(\partial_r^2+\frac{1}{r}\partial_r)
+\frac{m^2}{2r^2}+r^2\,,
\\
&&
H_2=-2p^+\partial_{p^+}^2 -\partial_{p^+}
+\frac{1}{2p^+}\Bigl(\kappa^2-\frac{1}{4}\Bigr) + p^+\,,
\end{eqnarray}
where $r$ is a radial variable $r\equiv |x^1+{\rm i}z|$ and
$\kappa\equiv m \pm 2s$.
Because $H_1$ and $H_2$ commute with each other we can decompose the wave
function as follows
$$
|\phi_{\pm s,m}\rangle=\phi^{(1)}_{\pm s,m}(r)
|\phi_{\pm s,m}^{(2)}(p^+)\rangle\,,
$$
where $\phi^{(1)}_{\pm s,m}(r)$ and
$|\phi_{\pm s,m}^{(2)}(p^+)\rangle$ are eigenvectors of $H_1$ and $H_2$
respectively.  Introducing instead of $p^+$ a new variable $y$, by
relation $y^2=p^+$ and rescaling wave function
$$
\phi_{\pm s,m}^{(1)}(r)=r^{-1/2}\tilde{\phi}_{\pm s,m}^{(1)}(r)\,,
$$
we get the following hamiltonians
\begin{eqnarray}
\label{ham1fin}
&&
H_1=
-\frac{1}{2}\partial_r^2+\frac{1}{2r^2}(m^2-\frac{1}{4})
+\frac{\omega_0^2}{2}r^2\,,
\\
\label{ham2fin}
&&
H_2=
-\frac{1}{2}\partial_y^2+\frac{1}{2y^2}(\kappa^2-\frac{1}{4})
+\frac{\omega_0^2}{2}y^2\,,
\end{eqnarray}
where $\omega_0\equiv\sqrt{2}$.
The eigenvalues of the hamiltonians (\ref{ham1fin}) and (\ref{ham2fin})
responsible for square integrable eigenvectors are well known and are
given by
$$
E_1=(2n_1+|m|+1)\omega_0\,,
\qquad
E_2=(2n_1+|m\pm 2s|+1)\omega_0\,,
$$
where $n_1,n_2=0,1,\ldots$.
Taking into account the relation (\ref{p0h1h2}) we get the following
eigenvalues of $\hat{P}^0$
\begin{equation}\label{enespe}
E=n_1+n_2+|\frac{m}{2}\pm s|+|\frac{m}{2}|+1
\end{equation}
A few comments are in order.
(i) Our energy spectrum (as usual) is discrete.
As compared to standard energy spectrum (see \cite{BRFR}) which depends on
two integers our energy spectrum (\ref{enespe})
depends on three integers $n_1$, $n_2$, $m$.
Such the difference is well know from analysis of quantum mechanical
energy spectrum of the standard three dimensional oscillator and is
related to a choice of coordinates in which energy spectrum is evaluated.
Normally, (see \cite{BRFR}) one uses global spherical coordinates while we
use Poincar\'e coordinates.  (ii) As usual for massless spin $s$ particle
in $AdS_4$ our energy spectrum is bounded from below by $E_{min}=s+1$.
(iii) There is 2-fold degeneracy of lowest energy $E_{min}$
related to the two
helicity states $|\phi_{\pm s}\rangle$.  (iv) for each helicity state
lowest energy has $(2s+1)$-fold  degeneracy, i.e. for $n_1=n_2=0$ there
exist $2s+1$ values of $m$ for which $E$ given in (\ref{enespe}) is equal
to $E_{min}$.  This degeneracy explains well known $so(3)$ symmetry of
lowest energy state.

\bigskip

{\bf Self-dual fields in AdS space-time.}
In this section we would like to discuss self-dual fields in AdS
space. In our knowledge they have not been  discussed previously in
literature.
As in Minkowski space the strengths of AdS (anti) self-dual fields satisfy
self-duality constraint
\begin{equation}\label{sfcst}
F_\pm^{\mu_1\ldots \mu_{d/2}}
=\pm\frac{1}{(d/2)!}
\frac{\epsilon^{\mu_1\ldots \mu_{d/2}\nu_1\ldots \nu_{d/2}}}{
\sqrt{|g|}}
F_{\pm\,\nu_1\ldots \nu_{d/2}}
\end{equation}
Because this constraint it is impossible to construct an appropriate
Lorentz covariant action without introducing auxiliary
fields\footnote{\scriptsize
Lorentz covariant formulations including auxiliary fields
are discussed in \cite{ber3},\cite{pst}.}.
The action for self-dual fields
can be most easily understood within the framework of light-cone gauge
formulation.  For definiteness let us restrict ourselves to the case of
six dimensional AdS space and to second rank antisymmetric tensor field.
First of all we would like to discuss formulation of self-dual fields in
Minkowski space which is most convenient for generalization to AdS space.
In light-cone gauge physical degrees of freedom of self-dual field in
Minkowski space are described by field $\phi^{IJ}$ which satisfies the
self-duality constraint $\phi^{IJ}=(1/2)\epsilon^{IJKL}\phi^{KL}$.  The
field $\phi^{IJ}$, which is the $so(4)$ tensor, can be decomposed into
$so(3)$ tensors $\phi^{ij}$ and $\phi^i$, $\phi^i\equiv\phi^{zi}$.
The self-duality constraint
tells us then that $\phi^{ij}$ is expressible in terms of $\phi^i$:
$\phi^{ij}=\epsilon^{ijk}\phi^k$.
Thus, if one wishes, in Minkowski
space the self-dual field can be described by unconstrained field
$\phi^i$ (or $\phi^{ij}$) \cite{ht}\footnote{\scriptsize
It is worth mentioning that formulation suggested in \cite{ht} (and its
generalization to curved space given in \cite{schwarz}) is
extremely appropriate to description of self-dual fields in AdS space.
The formulation given in \cite{ht} breaks Lorentz invariance
$so(d-1,1)$ to $so(d-2,1)$ invariance and this does not fit with manifest
Lorentz symmetry of Minkowski space metric.
On the other hand in AdS space it is the $so(d-2,1)$ symmetry that is
manifest symmetry of $AdS_d$ space metric
considered in Poincar\'e coordinates
(equivalently, the $so(d-2,1)$ is a manifest symmetry of AdS algebra
considered in conformal algebra notation).}.
Note that this form of description breaks manifest $so(4)$
(which is $so(d-2)$ for $d=6$) invariance to manifest $so(3)$ (which is
$so(d-3)$ for $d=6$) invariance.  On the other hand light-cone gauge
formalism in $AdS_d$ space respects only manifest $so(d-3)$ (see
\cite{metlc}). Therefore the description based on unconstrained
field $\phi^i$ is most appropriate to be generalized to AdS space.
Thus our aim is to find realization of AdS algebra generators
on the space of unconstrained field $\phi^i$. The realization of AdS
algebra on the space of physical fields found in (\cite{metlc}, formulas
(4.1)-(4.8) and (4.17)-(4.19)) is formulated in terms of spin operator
$M^{IJ}$ and operators $A$, $B$. The form of spin operator is fixed by
representation of $so(4)$ algebra we interested in.  To proceed we
introduce creation operator $\alpha^i$ and construct Fock space vector
\begin{equation}\label{2seldua}
|\phi\rangle\equiv \phi^i\alpha^i|0\rangle\,,
\qquad
\bar{\alpha}^i|0\rangle=0\,,
\qquad
[\bar{\alpha}^i,\alpha^j]=\delta^{ij}\,.
\end{equation}
For this form of realization the spin operators take the form
\begin{equation}\label{spiope}
M^{ij}=\alpha^i\bar{\alpha}^j-\alpha^j\bar{\alpha}^i\,,
\qquad
M^{zi}=\frac{1}{2}\epsilon^{ijk}M^{jk}\,.
\end{equation}
Note that it is second relation in (\ref{spiope}) that tells that we deal
with self-dual representation of $so(4)$ algebra. As to above mentioned
operators $A$ and $B$ they are fixed by defining equations found in
(\cite{metlc})
\begin{eqnarray}
\label{defcon1}
&&
2\{M^{zi},A\}-[[M^{zi},A],A]=0\,,
\\
\label{defcon2}
&&
[M^{zi},[M^{zj},A]]+\{M^{iL},M^{Lj}\}=-2\delta^{ij}B\,.
\end{eqnarray}
where operator $A$ is invariant under $so(d-3)$ spin rotations, i.e.
$[A,M^{ij}]=0$. From $[A,M^{ij}]=0$, the second relation in (\ref{spiope})
and equations (\ref{defcon1}) we find the equation $\{A,M^{zi}\}=0$ which
implies that $A=0$. By using (\ref{spiope}) it is easy to get then the
following relation
$$
\{M^{iL},M^{Lj}\}=-2\delta^{ij}M_{zl}^2\,,
\qquad
M_{zl}^2\equiv M^{zl}M^{zl}\,.
$$
From this relation and (\ref{defcon2}) we conclude that $B=M_{zi}^2$.
It is easy to check that in the case under consideration the $M_{zi}^2$
is diagonalized:  $M_{zi}^2|\phi\rangle=-2|\phi\rangle$. Due to
(\ref{spiope}) the $M_{zi}^2$ commutes with $M^{IJ}$ and therefore we can
put $B=-2$.  To summarize we have found
\begin{equation}\label{aib}
A=0\,,
\qquad
B=-2\,.
\end{equation}
Thus we found all entries of light-cone formulation. By substituting the
expressions for spin operators (\ref{spiope}) and operators $A$
and $B$ (\ref{aib}) in expressions (4.1)-(4.8) and (4.17)-(4.19) of
Ref.\cite{metlc} we find realization of AdS algebra generators on space of
AdS self-dual field $|\phi\rangle$ (\ref{2seldua}).
This provides complete description of self-dual field
$|\phi\rangle$ (\ref{2seldua}) in $AdS_6$ space.

Generalization to the case of arbitrary spin $s$ self-dual fields in
$AdS_6$ is straightforward. In this case we introduce
$$
|\phi\rangle
=\phi^{i_1\ldots i_s}\alpha^{i_1}\ldots \alpha^{i_s}|0\rangle\,,
$$
where $\phi^{i_1\ldots i_s}$ is totally symmetric traceless $so(3)$ tensor
field.  The spin operators take the form given in (\ref{spiope}) while for
operators $A$ and $B$ we get
$$
A=0\,,
\qquad
B=-s(s+1)\,.
$$
The light-cone gauge action for self-dual
field in AdS takes then the form
$$
S=\int d^dx\langle\partial^+\phi|(-\partial^-+P^-)|\phi\rangle\,,
\qquad
P^-=-\frac{\partial_I^2}{2\partial^+}\,,
$$
which coincides with the one in Minkowski space.
We think that this coincidence can be traced to the conformal
invariance of self-dual fields \cite{metsit3}.
Note that to describe anti self-dual fields one needs to
replace the second relation in (\ref{spiope}) by relation
$M^{zi}=-(1/2)\epsilon^{ijk}M^{jk}$. All remaining formulas above given
do not change their form.

\bigskip
{\bf Superstring theory and AdS higher spin massless fields theory.
Conjecture.} One of major motivations for our investigation of higher spin
massless fields theory in the AdS space is to seek a possible relation
between this theory and string theory. It seems rather attractive to
conjecture that string theory can be interpreted as resulting from some
kind of a spontaneous breakdown of higher spin symmetries. In \cite{metlc}
it has been conjectured that {\it superstrings could be considered
as the ones living at the boundary of $11$-dimensional AdS space
while their unbroken (symmetric) phase is realized as the theory
of higher spin massless fields living in this $AdS_{11}$ space}.
AdS theories are symmetric with respect to isometry algebra of
$AdS_d$ space which is $so(d-1,2)$. This algebra is not
realized however as symmetry algebra of string S-matrix . This implies that
$so(d-1,2)$ symmetry should be (spontaneously) broken.
Here we would like to demonstrate that if one restricts attention to
totally symmetric fields and make some mild assumptions about the
(spontaneously) broken form of AdS theory hamiltonian $P^-$ then some
interesting and non-trivial test of our conjecture can be carried out.
To this end we consider the $P^-$ for AdS totally symmetric fields
(\ref{totsym}). As was demonstrated in \cite{metlc} one has
\begin{equation}\label{pmads}
P^-=-\frac{\partial_I^2}{2\partial^+}
+\frac{1}{2z^2\partial^+}A\,,
\qquad
A
=-\frac{1}{2}M_{ij}^2+\frac{(d-4)(d-6)}{4}\,.
\end{equation}
where $M^{ij}$ is given in (\ref{dlc}). This $P^-$ can be rewritten as
\begin{equation}\label{pmcsm}
P^-=\frac{-\partial_i^2+M^2}{2\partial^+}\,,
\qquad
M^2
\equiv -\partial_z^2+\frac{1}{z^2}A\,.
\end{equation}
The operator $M^2$ in (\ref{pmcsm}) can be interpreted
as mass operator for a field propagating in $(d-1)$ dimensional Minkowski
spacetime while the $P^-$ (\ref{pmcsm}) can be
considered as hamiltonian of this field.
The operator $M^2$ given in (\ref{pmcsm}) has continuous spectrum. This
implies that AdS theories which are not supplemented by
(spontaneous) symmetry breaking lead to boundary theories which have
continuous mass spectrum. In order to get theory with discrete mass
spectrum one needs to break AdS symmetry. Now we have to make
assumption about term which breaks AdS symmetry. Our suggestion is to
consider the following ansatz for (spontaneously) broken $P_{s.b}^-$
\begin{equation}\label{pmspobro}
P_{s.b}^-
= P^- + \frac{\omega^2z^2-\bar{\omega}}{2\partial^+}\,,
\qquad
\omega\equiv \frac{1}{2\alpha^\prime}\,,
\qquad
\bar{\omega}\equiv \frac{d-1}{2\alpha^\prime}\,,
\end{equation}
where $P^-$ is given in (\ref{pmads}) and $\alpha^\prime$ is the universal
Regge slope parameter\footnote{\scriptsize
Here we discuss open AdS and
string theories. For the case of closed theories the $\omega$ and
$\bar{\omega}$ in (\ref{pmspobro}) should be multiplied by factor 2.}.
For this $P_{s.b}^-$ the mass operator takes the form
\begin{equation}\label{mspobro}
M_{s.b}^2 =
-\partial_z^2+\frac{1}{z^2}A+\omega^2 z^2 - \bar{\omega}\,.
\end{equation}
This $M_{s.b}^2$, in contrast to $M^2$ given in (\ref{pmcsm}),
has discrete spectrum. Let us evaluate this spectrum. To this end we
decompose  the  field $|\phi\rangle$, which transforms  in irreducible
representation of $so(d-2)$ algebra, into irreducible representations of
$so(d-3)$ subalgebra
$|\phi_{s^\prime}\rangle$
\begin{equation}\label{dec}
|\phi\rangle=\sum_{s^\prime=0}^s \oplus |\phi_{s^\prime}\rangle\,.
\end{equation}
Because of relation
$M_{ij}^2|\phi_{s^\prime}\rangle
=-2s^\prime(s^\prime+d-5)|\phi_{s^\prime}\rangle$
the operator $M_{s.b}^2$ for $|\phi_{s^\prime}\rangle$ takes
the form
\begin{equation}\label{mspobro2}
M_{s.b}^2
=-\partial_z^2+\frac{1}{z^2}(\nu^2-\frac{1}{4})
+\omega^2 z^2 -\bar{\omega}\,,
\qquad
\nu\equiv s^\prime+\frac{d-5}{2}\,.
\end{equation}
The spectrum of this operator is well known and is given by
\begin{equation}\label{mspobrospe1}
m_{s.b}^2=2\omega(\nu+2n+1)-\bar{\omega}\,,
\qquad
n=0,1,\ldots\,.
\end{equation}
From this it is seen that for leading term in decomposition (\ref{dec}),
i.e.  for $s^\prime=s$, and for $n=0$ the mass spectrum
is given by
\begin{equation}\label{mspobrospe2}
m_{s.b}^2=(s-1)/\alpha^\prime
\end{equation}
and this coincides exactly with the mass spectrum of massless and massive
string states belonging to leading Regge trajectory. Note that
overall normalization factor as well as additive
constant in r.h.s of relation (\ref{mspobrospe2}) are result of
the form of conjectured term in (\ref{pmspobro}) which breaks
(spontaneously) AdS symmetry.  What is extremely important is that the
dependence on $s$ in (\ref{mspobrospe2}) is essentially fixed by specific
form of AdS mass operator $A$ given in (\ref{pmads}).

By now due to \cite{vas1} it is known that to construct self-consistent
interaction of higher spin massless fields in $AdS_4$ it is necessary to
introduce, among other things, a infinite chain of massless totally
symmetric fields which consists of every spin just once \cite{vas1}. In
higher dimensions the totally symmetric fields should be supplemented by
mixed symmetry fields.  What is important however in higher dimensional
AdS space it is also necessary to introduce the same infinite chain of
massless totally symmetric fields, i.e. the one which consists of every
spin just once. It is this chain of totally symmetric fields that can be
found on leading Regge trajectory of string theory.

To summarize we have demonstrated that if (spontaneously) broken
hamiltonian $P^-$ takes the form given in (\ref{pmspobro}) then the
{\it leading components of AdS massless totally symmetric arbitrary spin $s$
states $|\phi_s\rangle$ (\ref{dec}) become massive string states belonging
to leading Regge trajectory}.

\newpage
{\small
%{\scriptsize
%{\tiny

}


\begin{thebibliography}{30}

\parskip-2pt

\bibitem{mal}
J. Maldacena,
\ATMP{2}{1998}{231};
S.S. Gubser, I.R. Klebanov, and A.M. Polyakov,
\PLB{428}{1998}{105};
E. Witten,
\ATMP{2}{1998}253

\bibitem{review}
O. Aharony, S.S. Gubser, J. Maldacena, H. Ooguri, and Y. Oz,
``Large N field theories, string theory and gravity'',
{\tt hep-th/9905111}


\bibitem{mt1}
R.R. Metsaev and A.A. Tseytlin,
\NPB{533}{1998}{109};
hep-th/9805028

\bibitem{krr}
R. Kallosh, J. Rahmfeld and A. Rajaraman,
{\it JHEP} {\bf 9809} (1998) 002,
hep-th/9805217

\bibitem{mt2}
R.R. Metsaev and A.A. Tseytlin,
\PLB{436}{1998}{281}
hep-th/9806095

\bibitem{k1}
R. Kallosh,
``Superconformal Actions in Killing Gauge'',
hep-th/9807206;
I.~Pesando,
%``A kappa fixed fixed type IIB superstring action on AdS(5) x   S(5),''
{\it JHEP} {\bf 11} (1998) 002,
{\tt hep-th/9808020};
R. Kallosh and J. Rahmfeld,
\PLB{443}{1998}{143};
hep-th/9808038;
R. Kallosh and A.A. Tseytlin,
\JHEP{10}{1998}{016};
hep-th/9808088

\bibitem{Dolan:1999pi}
L.~Dolan and M.~Langham,
%``Symmetric subgroups of gauged supergravities and AdS
%string theory vertex operators,''
{\it Mod. Phys. Lett.} {\bf A14} (1999)  517,
{\tt hep-th/9901030};
A.~Rajaraman and M.~Rozali,
``On the quantization of the GS string on $AdS_5
\times S^5$,''
{\tt hep-th/9902046};
D.~Berenstein and R.~G. Leigh,
``Superstring perturbation theory and
Ramond-Ramond backgrounds,''
{\tt hep-th/9904104}.

\bibitem{alter}
N. Berkovits, C. Vafa, E. Witten,
%``Conformal Field Theory of AdS Background with Ramond-Ramond Flux''
JHEP 9903 (1999) 018,
{\tt  hep-th/9902098};
M. Bershadsky, S. Zhukov, A. Vaintrob,
``$PSL(n|n)$ Sigma Model as a Conformal Field Theory''
{\tt hep-th/9902180};
hep-th/9905032;
N. Berkovits, M. Bershadsky, T. Hauer, S. Zhukov, B. Zwiebach,
``Superstring Theory on $AdS_2 x S^2$ as a Coset Supermanifold''
{\tt hep-th/9907200};
Jian-Ge Zhou,
\NPB{559}{1999}{92},
hep-th/9906013;

\bibitem{ber}
N. Berkovits,
``Quantization of the Type II Superstring in a Curved
Six-Dimensional Background''
hep-th/9908041;
N. Berkovits,
`Super-Poincare Covariant Quantization of the Superstring',
hep-th/0001035

\bibitem{twi}
P. Claus, M. Gunaydin, R. Kallosh, J. Rahmfeld, Y. Zunger,
%``Supertwistors as Quarks of $SU(2,2|4)$''
JHEP 9905 (1999) 019,
hep-th/9905112;
P. Claus, R. Kallosh, J. Rahmfeld,
\PLB{462}{1999}{285},
hep-th/9906195;
P. Claus, J. Rahmfeld, Y. Zunger
``A Simple Particle Action from a Twistor Parametrization of AdS(5)'',
hep-th/9906118;
Y. Zunger,
``Twistors and Actions on Coset Manifolds'',
hep-th/0001072;

\bibitem{GS5}
M. B. Green and J. H. Schwarz,
\PLB{122}{1983}{143}

\bibitem{GSB}
M. B. Green, J. H. Schwarz and L. Brink,
\NPB{219}{1983}{437}

\bibitem{metlc}
R.R. Metsaev, \NPB{563}{1999}{295},
hep-th/9906217

\bibitem{covequmot}
J.H. Schwarz,
\NPB{226}{1983}{269};
J.H. Schwarz and P.C. West,
\PLB{126}{1983}{301};
P.S. Howe and P.C. West,
\NPB{238}{1984}{181}

\bibitem{covact}
G. Dall'Agata, K. Lechner, D. Sorokin,
%``Covariant Actions for the Bosonic Sector of D=10 IIB Supergravity''
\CQG{14}{1997}{L195},
hep-th/9707044;
G. Dall'Agata, K. Lechner, M. Tonin,
\JHEP{9807}{1998}{017};
G.Arutyunov, S.Frolov
``Quadratic action for type IIB supergravity on $AdS_5\times S^5$''
hep-th/9811106

\bibitem{GS6}
M. B. Green and J.H. Schwarz,
\NPB{243}{1984}{475}

\bibitem{metiib}
R.R. Metsaev, \PLB{468}{1999}{65}, hep-th/9908114

\bibitem{metsit3}
R.R. Metsaev,
\MPLA{10}{1995}{1719}

\bibitem{krn}
H.J. Kim, L.J. Romans and P. van Nieuwenhuizen,
\PRD{32}{1985}{389}


\bibitem{peslc}
I. Pesando
``On the Gauge Fixing of the $\kappa$ Symmetry on AdS and Flat
Background: the Lightcone Action for the Type IIb String on
AdS(5) X S(5)'',
hep-th/9912284

\bibitem{BRFR}
P. Breitenlohner and D.Z. Freedman
\AP{144}{1982}{249}

\bibitem{ber3}
N. Berkovits,
\PLB{395}{1997}{28}, hep-th/9610134.

\bibitem{pst}
P. Pasti, D. Sorokin and M. Tonin,
\PRD{55}{1997}{6292},
hep-th/9611100


\bibitem{ht}
M. Henneaux and C. Teitelboim,
\PLB{206}{1988}{650}

\bibitem{schwarz}
J.H. Schwarz,
\PLB{395}{1997}{191},
hep/9701008



\bibitem{vas1}
M.A. Vasiliev,
\PLB{243}{1990}{378};
\CQG{8}{1991}{1387}

\end{thebibliography}
\end{document}